\documentclass[%
superscriptaddress,
reprint,
 amsmath,amssymb,
 aps,
 pra,
 ]{revtex4-2}

\usepackage{color}
\usepackage{graphicx}
\usepackage{dcolumn}
\usepackage{bm}
\usepackage[caption=false]{subfig}
\usepackage[pdftex, pdfborderstyle={/S/U/W 0}]{hyperref} 
\newcommand{\change}[1]{{#1}}

\begin{document}

\title{Impacts of Noise and Structure on Quantum Information Encoded in a Quantum Memory}

\author{Matthew Otten}
\email[Correspondence and Present Address: ]{mjotten@hrl.com, HRL Laboratories, LLC, 3011 Malibu Canyon Road, Malibu, California 90265}
\affiliation{%
  Nanoscience and Technology, Argonne National Laboratory, Lemont, Illinois 60439
}%
\author{Keshav Kapoor}

\author{A. Bar\i\c{s} \"Ozg\"uler}
\author{Eric T. Holland}
\altaffiliation{Present Address: Quantum R\&D Center, Keysight Technologies Inc., 1 Broadway, Cambridge, Massachusetts 02142}
\author{James B. Kowalkowski}
\affiliation{%
  Fermi National Accelerator Laboratory, Batavia, Illinois 60510
}%
\author{Yuri Alexeev}
\affiliation{%
  Computation Science Division, Argonne National Laboratory, Lemont, Illinois 60439
}%
\author{Adam L. Lyon}
\affiliation{%
  Fermi National Accelerator Laboratory, Batavia, Illinois 60510
}%

\date{November 25, 2020}

\begin{abstract}
As larger, higher-quality quantum devices are built and demonstrated in quantum
information applications, such as quantum computation and quantum communication,
the need for high-quality quantum memories to store quantum states becomes ever
more pressing. Future quantum devices likely will use a
variety of physical hardware, some being used primarily for  processing of quantum information and others for storage. Here, we study the
correlation of the structure of quantum information with physical noise models
of various possible quantum memory implementations. Through numerical
simulation of different noise models and approximate analytical formulas applied
to a variety of interesting quantum states, we provide comparisons between
quantum hardware with different structure, including both qubit- and qudit-based
quantum memories. Our findings point to simple, experimentally relevant formulas
for the relative lifetimes of quantum information in different quantum memories
and have relevance to the design of hybrid quantum devices.
\end{abstract}

\maketitle


\section{\label{sec:intro}Introduction}
In many quantum information science technologies, quantum memories, which store
quantum states until they are required, are an integral part of the overall
architecture~\cite{lvovsky_optical_2009,dennis_topological_2002}. For instance,
in quantum computing applications, the ability to store a state between
computations is necessary for increasing the overall fidelity of a quantum
algorithm~\cite{ladd2010quantum}. In quantum communication protocols,
quantum memories make up a key part of several quantum repeater
designs~\cite{duan2001long,briegel1998quantum}. Quantum error correction offers
the ability to create an arbitrarily long-lived quantum memory, with the overhead
depending on the size and complexity of the quantum
hardware~\cite{gottesman1997stabilizer}, through various protocols such as
surface codes~\cite{fowler2012surface}. In the noisy intermediate-scale
quantum (NISQ) era, however, the hardware overhead from performing quantum error
correction prevents its use~\cite{preskill2018quantum}. Nevertheless, even without error
correction, today's quantum hardware is already performing impressive
demonstrations, such as quantum supremacy~\cite{arute2019quantum}, quantum
calculations in quantum chemistry~\cite{kandala-nature-2017,omalley-prx-2016,peruzzo-ncomms-2014},
quantum simulations of many-body physics~\cite{Edwards2010, Kim2011, Zhang2017, Kyriienko2018, Schauss2018, Quantum2020},
quantum dynamics~\cite{otten2019noise,chiesa2019quantum}, quantum optimization~\cite{venturelli2015quantum, bengtsson2020improved}, quantum machine
learning~\cite{otten2020quantum,havlivcek2019supervised}, quantum internet~\cite{Awschalom, valivarthi2020teleportation}, and quantum networking
demonstrations over long distances using
satellites~\cite{yin2020entanglement, Gundo}, optical
fibers~\cite{krutyanskiy2019light, yu2020entanglement}, and photonic quantum repeaters~\cite{hasegawa2019experimental}. Going beyond
these impressive, albeit small-scale, demonstrations will require high-quality
quantum memories. Quantum memories can be made out of many candidate hardware
platforms, including photonics~\cite{wang2020integrated}, superconducting
cavities~\cite{ofek2016extending},  superconducting qubits~\cite{neeley2008process},
vacancy centers~\cite{Lai2018},
trapped ions~\cite{monz_realization_2009}, and
silicon quantum dots~\cite{Sigillito2019}.
Each platform offers various benefits and drawbacks, for instance, in coherence
times, fabrication difficulty, and interoperability.

Here, we study the performance of storing a variety of quantum states in various
quantum memories with differing noise models, exploring the correlation of the
structure of the stored quantum state with the structure of the noise model.
We focus primarily on the difference of storing quantum information in
qubit-based systems, where the state is stored in a possibly entangled register
of qubits, to many-level qudit-based systems, where the state can be stored in
one single quantum system with many levels. We provide extensive numerical
calculations of such systems under amplitude damping ($T_1$) and dephasing
($T_2^*$) noise models, as well as simple analytic formulas for predicting the
coherence requirements for the different systems and noise models to have the
same memory performance. Our results point to qudit-based systems as being
viable candidates for high-quality quantum memories, given the ability to
engineer extremely coherent superconducting cavity
systems~\cite{reagor_reaching_2013,reshitnyk_3d_2016, xie2018compact, Romanenko_3d}.

\section{Theoretical Methods}
\begin{figure}
  \includegraphics[width=\columnwidth]{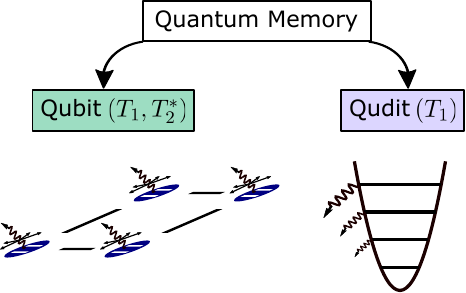}
  \caption{Quantum memories can be made out of a variety of quantum systems;
    here, we schematically show the noise properties of two quantum memories.
    Left: A quantum memory comprising qubits subject to amplitude damping (also
    called $T_1$, represented by the waves flying out of the qubits) and pure
    dephasing (also called $T_2^*$, denoted by the arrows pointing in many
    directions). Each qubit adds two noise channels, each with the same
    strength. Right: A quantum memory comprising a single qudit subject to
    amplitude damping. The noise grows as the number of levels increases.}
  \label{fig:schematic}
\end{figure}

In this section, we describe the considered noise models, how we encode quantum states, and the particular states we considered. Furthermore, to complement our numerical analysis we introduce an analytical study based on approximating the Lindblad master equation with a non-Hermitian formalism.

\subsection{Noise Models}
We consider the evolution of a quantum state under a noisy channel using the
Lindblad master equation,
\begin{equation}\label{eqn:lme}
  \frac{d \rho (t) }{d t} = \sum_{i}\gamma_i  L(C_i)[\rho(t)],
\end{equation}
where $\rho$ is the density matrix of the system, $\gamma_i$ is a noise rate,
$L(C_i)[\rho(t)] = \left(C_i\rho(t) C_i^{\dagger} -
  \frac{1}{2}\left\{C_i^{\dagger}C_i,\rho(t)\right\}\right)$ is a Lindblad
superoperator, and $C_i$ are operators
representing various noise processes.
\change{The Lindblad master equation is one of the standard approaches
for studying Markovian open quantum systems~\cite{carmichael2009open,haroche2006exploring}
and has been used to
study many physical systems, such as superconducting qubits~\cite{gambetta2008quantum}
and quantum dots~\cite{otten_origins_2016,otten-prb-2015}.}
In this work, we
consider only incoherent evolution of the system under various noise processes; \change{thus, the Hamiltonian is chosen to be zero}.
For any noise model, if
$\gamma_i=0$, the state will be maintained perfectly in the quantum system for
all time. We study both amplitude damping noise ($T_1$), where the $C_i$ are \change{annihilation} operators, and dephasing ($T_2^*$) noise,
where the $C_i$ are number operators. These are dominant noise
sources on a variety of NISQ hardware previously used in a
variety of
studies~\cite{macquarrie_cooling_2017,otten-prb-2015, Remizov2019, Paz-Silva2019, Sung2020}. Furthermore, we study the difference between encoding the quantum
information into qubit-based hardware, where the quantum state is represented
by a possibly entangled register of multiple two-level systems, and into
qudit-based hardware, where the quantum state is represented by a possibly
entangled register of $d$-level systems. This is shown schematically in
Fig.~\ref{fig:schematic}. Although there are many possible
combinations of amplitude damping, dephasing, and number of levels of the
quantum hardware, we focus primarily on two specific combinations: qubit-based
systems with both amplitude damping and dephasing noise, which is paradigmatic
of superconducting transmon qubits~\cite{Krantz}, and a single many-level qudit, with
only amplitude damping noise, which is paradigmatic of superconducting
cavities\cite{reagor_reaching_2013, Reagor2016}. The specific Lindblad master equation for the qubit-based noise
model with both amplitude damping and dephasing noise is then
\begin{equation}\label{eqn:qb_lme}
  \frac{d \rho(t)}{d t} = \sum^{n_q}_{i}\gamma(L(\sigma_i)[\rho(t)] + L(\sigma_i^\dagger \sigma_i)[\rho(t)]),
\end{equation}
where $n_q$ is the number of qubits necessary to store the quantum state,
$\sigma_i$ is the \change{annihilation} operator for a two-level system, and we have
assumed that all  the noise rates for all qubits, $\gamma$, are the same.
The specific Lindblad master equation for the single qudit-based noise
model with only amplitude damping is
\begin{equation}\label{eqn:qd_lme}
  \frac{d \rho(t)}{d t} = \gamma L(b)[\rho(t)],
\end{equation}
where $b$ is the \change{annihilation} operator for a $d$-level system large enough to
store the full quantum state. \change{As an example, for a qudit with $d=4$, we use the matrix
\begin{equation}
   b =  \begin{bmatrix}
0 & \sqrt{1} & 0 & 0\\
0 & 0 & \sqrt{2} & 0 \\
0 & 0 & 0 & \sqrt{3}\\
0 & 0 & 0 & 0 \\
\end{bmatrix}.
\end{equation}
Larger qudits are similarly defined, with the matrix being size $d\times d$ and
$b_{i,i+1} = \sqrt{i}$ for every row but the last, which has no nonzero elements.}
In the case of storing the same state in both a
qubit-based system and a single qudit-based system, \change{$d=2^{n_q}$}. Although we will
focus primarily on these two models, our analysis is easily generalized to other
systems (as we later discuss), such as qubit-based systems with only
dephasing, which could represent ion-trap quantum memories~\cite{Harty2014}, or a system
with \change{$d<2^{n_q}$}, which would represent an array of possibly entangled qudits.

To quantify the performance of these noise models, we study how well
various quantum states are preserved after evolution under the different
Lindblad master equations, such as Eq.~\eqref{eqn:qb_lme} and
Eq.~\eqref{eqn:qd_lme}. We calculate the fidelity of the state as it
evolves under the noisy channel, and we directly compare the time for various noise
models to reach a fidelity target of $\mathcal{F}_t$. The ratio of times
$\frac{t_a}{t_b}$ for some noise model $a$ and a different noise model $b$ to reach the
fidelity target provides a direct comparison of the two noise models' performance
as quantum memories. Furthermore, since we study only incoherent evolution under
the Lindblad master equation with a single parameter $\gamma$, the ratio
$\frac{t_a}{t_b}$ directly provides the needed scaling in noise rates between
the two models to have an equal performance for the specified target fidelity.
\change{This allows us to simply use
$\gamma=1$ in arbitrary units for all simulations and then freely rescale
the units of $\gamma$ in post-processing, resulting in only needing one simulation
for each noise model for each state. Finding the ratio $\frac{t_a}{t_b}$ where
the two noise models both reach the target fidelity, $\mathcal{F}_t$, then directly
provides the relative scale of the two noise models $\gamma$, represented in the same units.}
\change{We define the fidelity between wavefunctions $|\psi\rangle$ and
$|\phi\rangle$ as $F=|\langle\psi|\phi\rangle|^2$.}
We
choose a fidelity target of $\mathcal{F}_t=0.75$ for our numerical studies because of its relevance in distinguishing multipartite entangled
states~\cite{acin_classification_2001}, but the results are not sensitive to this choice, as we show in our approximate analysis and in
numerical studies in Appendix~\ref{ap:ftar}. We use the open quantum systems solver
QuaC~\cite{QuaC:17} to perform numerical integration of the various Lindblad
master equations. \change{The QuaC code numerically propagates the Lindblad master equation
using an explicit Runge-Kutta time stepping scheme with an adaptively selected time step size~\cite{arfken1999mathematical}.
The density matrix is vectorized and stored as a dense column vector. The Lindbladians are stored
in a sparse matrix format to reduce the memory overhead.}

\subsection{Encoding Quantum States}
To evaluate the performance of the quantum systems as quantum memories, we study
the storage of a large variety of interesting quantum states. A generic quantum
state can be written as
\begin{equation}
  |\psi\rangle = \sum_j \alpha_j |j\rangle,
\end{equation}
where $\alpha_j$ is the amplitude of state $|j\rangle$.
\change{We restrict our study to only pure states.}
To map an arbitrary
state to a specific quantum system, whether it is based on qubits or qudits, we
map the amplitudes to various states of the quantum system. In the
case of qubit systems, we map the amplitude $\alpha_j$ to the qubit state
represented by the binary representation of the integer $j$. For example,
$\alpha_5$ is mapped to the qubit state $|101\rangle$. More generally, the
state $j$ is mapped to the $d$-nary representation of the integer $j$.
The specific states we study in this work include
multipartite entangled states such as the \change{Greenberger–Horne–Zeilinger (GHZ)}
and W states~\cite{dur2000three},
which find use in various quantum sensing protocols~\cite{degen_quantum_2017},
the equal superposition state, which is used in the initialization of many
quantum algorithms~\cite{yu_generic_2017} such as the Deutsch--Jozsa
algorithm~\cite{deutsch1992rapid}, Fock states, the coherent state
\change{(abbreviated `Coh.' in figures)} which
sees use in quantum algorithms for machine learning~\cite{otten2020quantum} and simulations of
quantum field theories~\cite{Macridin2018}, the ground and first excited state of H$_2$, H$_4$,
LiH, and H$_2$O, which are the result of quantum chemistry algorithms such as the
variational quantum
eigensolver \change{(VQE)}~\cite{colless_computation_2018,peruzzo-ncomms-2014}, the result of
running the quantum approximate optimization algorithm \change{(QAOA)} on the MaxCut
problem~\cite{farhi2014quantum,shaydulin2019network}, states with random
amplitudes on each state $|j\rangle$ \change{(abbreviated `Arb.' in figures)},
and the tensor product of random qubit states
\change{(abbreviated `Unent.' in figures)}. Further
description of these states can be found in Appendix~\ref{ap:states}.

\subsection{Non-Hermitian Analysis}
In addition to our numerical studies, we  provide an analysis based
on the approximation of the Lindblad master equation, Eq.~\eqref{eqn:lme}, with
a non-Hermitian formalism. Such formalism has been previously used to study the
contributions of both amplitude damping~\cite{otten_origins_2016} and
dephasing~\cite{cortes2020non} and can be identified as the first stage of the
Monte Carlo wavefunction approach before a stochastic
collapse~\cite{molmer1993monte}. In this approach, instead of studying the time
evolution of the density matrix, we study the evolution of the
wavefunction under a non-Hermitian Hamiltonian,
\begin{equation}\label{eqn:nh}
  \frac{d |\psi(t)\rangle}{dt} =  \sum_i \frac{-\gamma_i}{2} C_i^\dagger C_i |\psi(t)\rangle,
\end{equation}
where $|\psi(t)\rangle$ is the wavefunction of the state, $\gamma_i$ and
$C_i$ are the same as in Eq.~\eqref{eqn:lme}, and we have used units such that
$\hbar=1$. Note that, because we are using a
non-Hermitian formalism, there is no imaginary unit, $i$, in this equation.
Evolution of this non-Hermitian system leads to loss of the
overall norm of the wavefunction, which is the primary source of error, since the
norm is never recovered~\cite{cortes2020non}. This formalism is a
powerful tool, however, allowing for approximate analysis of the evolution of the fidelity
for arbitrary states and exact solutions for a small handful of states. For
example, the evolution of the fidelity of the qubit-based noise model with both
amplitude damping and dephasing under the non-Hermitian approximation is
\begin{equation}\label{eqn:qb_fid}
  \sqrt{F(t)} = \sum_{j}|\alpha_j|^2 e^{-\gamma w(j)t},
\end{equation}
where the initial state $|\psi(0)\rangle=\sum_j \alpha_j |j\rangle$ for
eigenstates $|j\rangle$  and $w(j)$ is the Hamming weight of the integer
$j$. The evolution of the fidelity of the qudit-based noise model with only
amplitude damping under the non-Hermitian approximation is
\begin{equation}\label{eqn:qd_fid}
  \sqrt{F(t)} = \sum_{j}|\alpha_j|^2 e^{-\frac{\gamma}{2} jt}.
\end{equation}
Full derivations of both  equations can be found in Appendix \ref{ap:A}.
We use these solutions  both to predict how the fidelity will evolve for systems
larger than can be reasonably simulated and to provide intuition and a simple
approximate formula for predicting relative performance between various noise
models. \change{We note that the non-Hermitian formalism is used
only as a tool in the analytic derivations. All numerically simulated
data is produced using the full Lindblad equation.}

\section{Results}

In this section, we compare qubit and qudit quantum memory architectures with their associated noise models. We concentrate at first on the GHZ initial state, showing numerical results and the analytical prediction for the ratio of times for the state to reach the target fidelity in the two quantum memory systems. We then expand the explanation to include other initial quantum states of interest. For many quantum states, the surprisingly simple analytical approximation is in close agreement with the numerical simulation.

\subsection{GHZ State}

The GHZ state is one of the primary genuine multipartite, maximally entangled
states~\cite{dur2000three} and is one of the canonical states used for
entanglement-enhanced quantum sensing~\cite{degen_quantum_2017}.
\change{We choose to initially focus on the GHZ state as an explicit example of carrying
out all of the steps of our analysis. It has many appealing features as an initial demonstration,
including a simple description when mapped to both qubit and qudit memories which aids in the
analytic derivation.}
For a
collection of $n_q$ qubits, the GHZ state is defined as
\begin{equation}\label{eq:ghz_state}
  |\mathrm{GHZ}\rangle = \frac{|0\rangle^{\otimes n_q} + |1\rangle^{\otimes n_q}}{\sqrt{2}}.
\end{equation}
That is, the GHZ state is a superposition of all  the qubits in the
$|0\rangle$ state with all  the qubits in the $|1\rangle$ state. When using a
qubit-based hardware, the mapping of the GHZ state is directly given by the
definition of state, Eq.~\eqref{eq:ghz_state}. Mapping to the qudit state by
the construction mentioned above gives $\frac{1}{\sqrt{2}}\big(|0\rangle +
|2^{n_q}-1\rangle\big)$. Given that this state has only two amplitudes, it is
 simple to apply the non-Hermitian analysis described above.
The fidelity for the qubit-based quantum memory with both amplitude damping and
dephasing, following Eq.~\eqref{eqn:qb_fid}, is approximately
\begin{equation}\label{eqn:qb_ghz}
  \sqrt{F(t)} = \frac{1}{2} \big(1 + e^{-\gamma n_q t}\big).
\end{equation}
The fidelity for the qudit-based quantum memory with only amplitude damping, on the
other hand, following Eq.~\eqref{eqn:qd_fid}, is approximately
\begin{equation}\label{eqn:qd_ghz}
  \sqrt{F(t)} = \frac{1}{2} \big(1 +e^{-\gamma (2^{n_q} - 1) t}\big).
\end{equation}
Comparing the two quantum memory architectures, the qubit-based
(Eq.~\eqref{eqn:qb_ghz}) and the qudit-based (Eq.~\eqref{eqn:qd_ghz}), one
immediately sees that the qudit-based quantum memory will perform
exponentially worse as the number of qubits grows, since its effective decay
rate is exponentially larger. This also follows the intuition behind the two
different noise models, shown schematically in Fig.~\ref{fig:schematic}. For the
qubit-based quantum memory, each additional qubit adds another independent
channel of amplitude damping and dephasing with rate $\gamma$. In contrast, for
the qudit-based quantum memory, encoding an additional qubit's worth of
information requires doubling $d$, the total size of
the qudit. Each additional level of the qudit decays faster than the previous.
Doubling the number of levels effectively doubles the decay rate. In the GHZ
state, where only the lowest level ($|0\rangle$) and the highest level
($|2^{n_q}-1\rangle$) are included, this effect is clearly demonstrated. To quantify the
difference in performance, we calculate the ratio of the qubit-based
quantum memory to reach a target fidelity $\mathcal{F}_t$ (denoted  $t_b$) and the time
for the qudit-based quantum memory to reach the same (denoted  $t_d$). Through
simple algebra, we find
\begin{equation}\label{eqn:ghz_ratio}
  \frac{t_b}{t_d} = \frac{2^{n_q}-1}{2n_q},
\end{equation}
assuming that both quantum memories have the same noise rate $\gamma$. For the
GHZ state under the non-Hermitian approximation, this ratio is independent of
the target fidelity. As discussed in the Methods section, this ratio provides a
direct, quantifiable comparison of the two quantum memories. It can also be
interpreted as the  decrease in noise necessary to make the qudit-based
quantum memory perform as well as the qubit-based quantum memory.

\begin{figure}
  \includegraphics[width=\columnwidth]{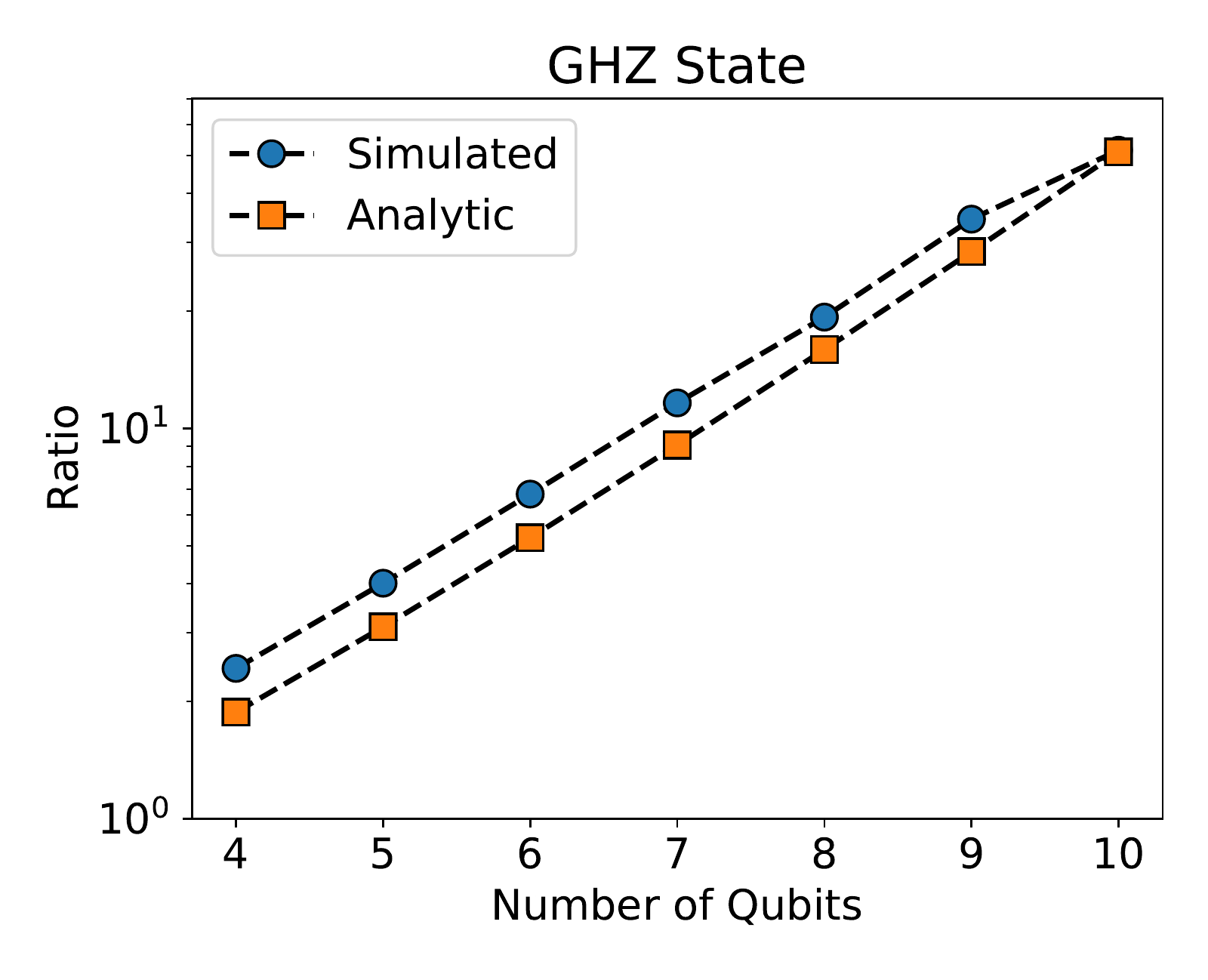}
  \caption{Scaling ratio for the qudit-based quantum memory to perform as well
    as the qubit-based quantum memory versus the number of qubits for the GHZ
    state. The ratio grows exponentially as the number of qubits grows.}
  \label{fig:ghz}
\end{figure}

Figure~\ref{fig:ghz}
shows the analytic ratio of Eq.~\eqref{eqn:ghz_ratio}, as well as the
numerically simulated ratio using the full Lindblad master equations of the
qubit-based quantum memory (Eq.~\eqref{eqn:qb_lme}) and the qudit-based quantum
memory (Eq.~\eqref{eqn:qd_lme}) with target fidelity $\mathcal{F}_t=0.75$. The
ratio needed for both the analytic formula and the numerical simulations grows
exponentially with the number of qubits, as expected. The analytic formula
slightly underestimates the scaling ratio needed; this is to be expected since
the inclusion of dephasing in the non-Hermitian formalism has been shown to
significantly increase the approximation error~\cite{cortes2020non}.

\subsection{Predicted Ratio}
The GHZ state, a superposition of only two states, is
the simplest state in this study. Generally, the coefficients $\alpha_j$ can take
any (normalized) set of values. To provide an approximate ratio for an arbitrary
state, we start with the general expression of the fidelity using the solution to the
Schr\"{o}dinger equation,
Eq.~\eqref{eqn:sch_sol},
\begin{equation}\label{eqn:sqrt_fid}
  \sqrt{F(t)} = \Big|\langle \psi(0) | e^{-\sum_j \frac{\gamma_j}{2} C_j^\dagger C_j } | \psi(0) \rangle\Big|.
\end{equation}
Equation~\eqref{eqn:sqrt_fid} is valid for any initial condition, $|\psi(0)\rangle$ and
any system with any noise operators, $C_i$. To obtain an approximate solution
for this equation, we expand the exponential for some target fidelity, $\mathcal{F}_t$,
\begin{equation}\label{eqn:sqrt_fid_expand}
  \sqrt{\mathcal{F}_t} = \sum_k \sum_j \frac{(-\gamma_j t)^k}{2^k k!} \langle  m^k_j \rangle,
\end{equation}
where  $\langle  m_j^k \rangle =  \big|\langle \psi(0) | \big(C_j^\dagger C_j\big)^k | \psi(0)
\rangle\big|$ are the moments of the operator $C_j^\dagger C_j$ for the state $|\psi(0)\rangle$.
This is valid for any state, $|\psi(0)\rangle$, and set of noise operators,
$C_i$. For example, using a qudit-based quantum memory with only amplitude
damping, we have only one noise operator, $b$, whose moments are simply the
moments of the number operator for a qudit, $\langle  n_d^k \rangle$.
We can rewrite the fidelity expansion of
Eq.~\eqref{eqn:sqrt_fid_expand} as
\begin{equation}\label{eqn:qd_exp_n}
  \sqrt{\mathcal{F}_t} = \sum_k \frac{(-\gamma t_d)^k}{2^k k!} \langle n_d^k \rangle.
\end{equation}
A similar expansion can be obtained for qubit-based quantum memories,
\begin{equation}\label{eqn:qb_exp_n}
  \sqrt{\mathcal{F}_t} = \sum_k \frac{(-\gamma t_b)^k}{k!} \langle n_b^k \rangle,
\end{equation}
where, analogous to the qudit-based quantum memory,  $\langle n_b^k \rangle =
\big\langle \psi(0) \big| \big(\sum_i \sigma_i^\dagger \sigma_i\big)^k \big|
\psi(0)\big\rangle$. The
difference in the factor of $2^k$ is due to the inclusion of dephasing in the
qubit-based quantum memory noise model. Since we have chosen a specific target
fidelity, $\mathcal{F}_t$, these two equations, Eq.~\eqref{eqn:qd_exp_n} and
Eq.~\eqref{eqn:qb_exp_n}, can be equated, and an approximation for the ratio of
the times to reach the target fidelity can be obtained by truncating the sum to
first order:
\begin{equation}\label{eqn:ratio_prediction}
  \frac{t_b}{t_d} \approx \frac{\langle n_d \rangle}{2\langle n_b \rangle}.
\end{equation}
We have shown here that the ratio of the times for the two quantum memories to
reach the target fidelity can simply be approximated as the ratio of the average
number of excitations in the qudit-based quantum memory compared with double the
average number of excitations in the qubit-based quantum memory. To first order,
the ratio does not depend on the target fidelity. Higher-order
truncations will depend on the target fidelity.
\change{For example, keeping terms to second order gives the following equation
\begin{equation} \label{eqn:ratio_pred_2nd}
  \frac{t_b}{t_d} \approx \frac{\langle n_d \rangle}{2\langle n_b \rangle} + \frac{\gamma}{\langle n_b\rangle t_d} \Big(t_b^2\langle n_b^2\rangle - t_d^2 \langle n_d^2\rangle\Big)
\end{equation}
which is a transcendental equation, that,
in general, is not solvable in closed form. However, specifying a specific target fidelity,
$\mathcal{F}_t$, and solving truncated forms of eqns.~\eqref{eqn:qd_exp_n} and~\eqref{eqn:qb_exp_n}
via, e.g, a root-finding technique is possible.
From this second-order expansion, we can see that the
our simpler formula, eq.~\eqref{eqn:ratio_prediction}, is accurate to first-order in the decay rate,
$\gamma$ and has terms that depend on the second moments of the specific quantum state.
}

\change{The simple first order equation, eq~\eqref{eqn:ratio_prediction}},
can be intuitively understood as representing
the correlation of the quantum state and the noise
models that were used to describe the quantum memories. For example, in the
qubit-based quantum memory, each excitation (i.e., the qubit in the $|1\rangle$
state) of an individual qubit contributes
to the overall noise by being subject to an amplitude damping and dephasing noise
channel. However, if the qubit is not excited (i.e., the qubit is in the
$|0\rangle$ state), there is no contribution to the
overall noise. A superposition of being excited ($|1\rangle$) and not being
excited ($|0\rangle$) would contribute only partially to the total amount of
noise. Therefore, under this intuitive argument, we can say that the total
amount of noise goes as $2\gamma \langle n_b \rangle$. Similar arguments can be
made for the qudit-based noise model, leading  to the total amount of noise
being, intuitively, $\gamma \langle n_d \rangle$. The ratio of the total amount
of noise between the two quantum memories should, then, give some insight into
their relative performance. As derived in Eq.~\eqref{eqn:ratio_prediction}, this
ratio is approximately the ratio of times to reach any target fidelity.

\begin{figure}
  \includegraphics[width=\columnwidth]{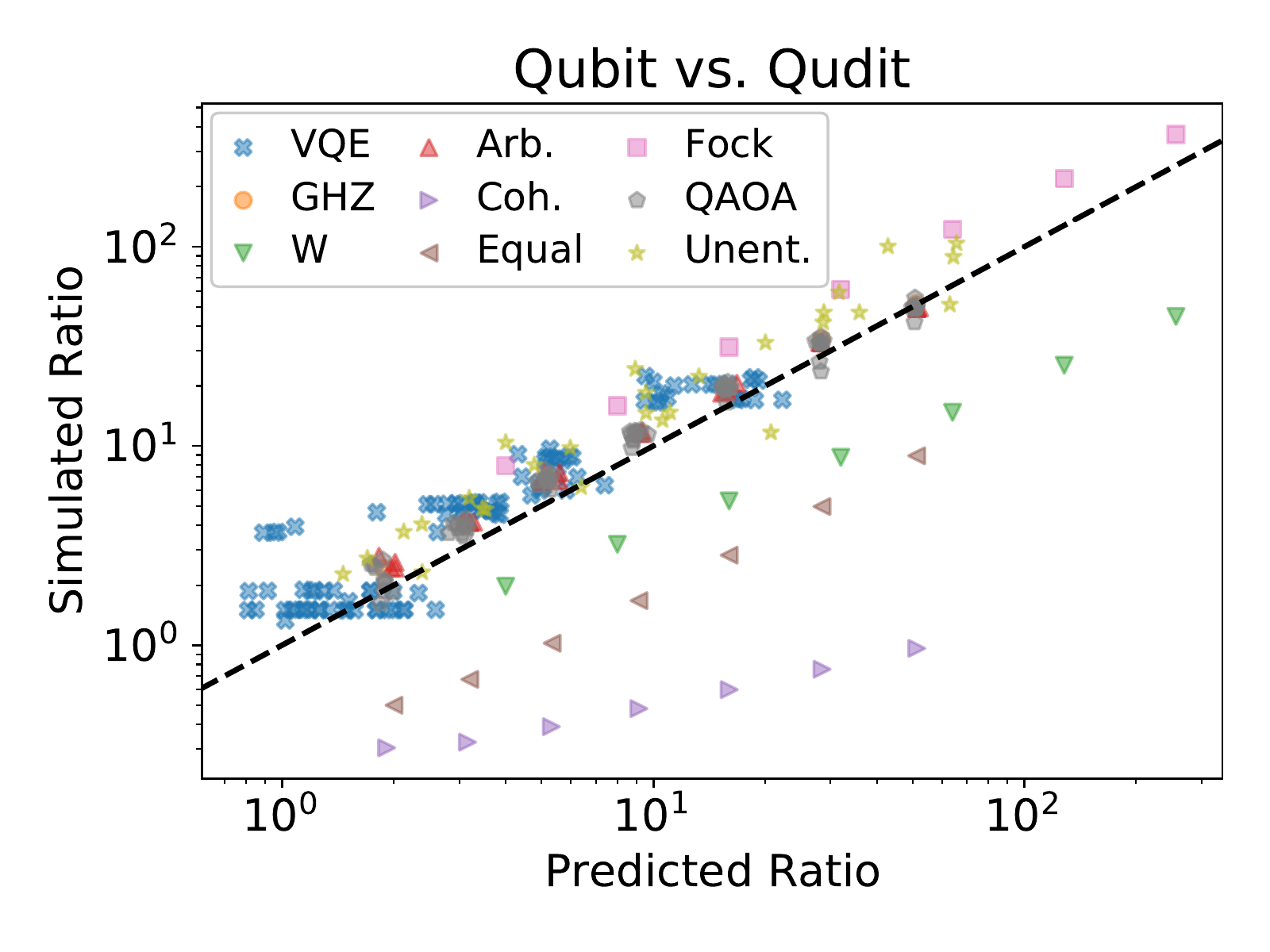}
  \caption{Comparison of the numerically simulated and analytically predicted scaling ratios between a
    qubit-based quantum memory with both amplitude damping and dephasing and a
    qudit-based quantum memory with only amplitude damping, for a wide of variety
    of interesting quantum states. The dashed black line is $y=x$; the closer
    the points are to this line, the better the prediction. }
  \label{fig:ratio_pred}
\end{figure}

Figure~\ref{fig:ratio_pred} plots the ratio from data simulated by using the
Lindblad master equations for the qubit-based (Eq.~\eqref{eqn:qb_lme}) and the
qudit-based (Eq.~\eqref{eqn:qd_lme}) quantum memories over a wide range of
interesting quantum states and sizes. The set of states is discussed in the
Theoretical Methods section. System sizes range from $2^4$ to $2^{15}$. The line $y=x$ is also
indicated on the plot; the closer the points are to this line, the better
the prediction of the approximate formula, Eq.~\eqref{eqn:ratio_prediction}. The
simulated and predicted ratios are highly correlated for most of the states. The coherent
and equal superposition states stray far from the line $y=x$.
\change{Notably, these two states are the states in which the non-Hermitian dynamics
diverges from the full dynamics of the Linblad master equation, as
we show in Appendix~\ref{ap:comparison}. The failure of our predicted ratio is thus because
of the failure of the non-Hermitian approximation. For the coherent state, specifically,
this is perhaps to be expected, as the full dissipative dynamics have a simple analytically
derivable form~\cite{haroche2006exploring}, which is very different than our approximate non-Hermitian form.}

This level of agreement is  remarkable, given that the derivation of the
approximate ratio
involves multiple levels
of approximation, \change{including both the
non-Hermitian approximation used to derive the fidelity equations,
eq.~\eqref{eqn:qd_exp_n} and~\eqref{eqn:qb_exp_n}, as well as their truncation
to first order. The contribution of the truncation was discussed above. To understand the error
from the non-Hermitian formalism used to derive the fidelity equations, we
compare the full Lindblad dynamics and the non-Hermitian dynamics in Appendix~\ref{ap:comparison}
for the various states studied.}
\change{Our approximate ratio} is surprisingly simple and comprises quantities
that can be easily measured in an experimental setting for an arbitrary, unknown
quantum state.
The use of such a formula is not limited to just the two quantum
memories studied here; it is easily generalized to other architectures and can
help inform possible strategies for extending the lifetimes of quantum
information in quantum memories.

\section{Discussion}
\begin{figure}
  \includegraphics[width=\columnwidth]{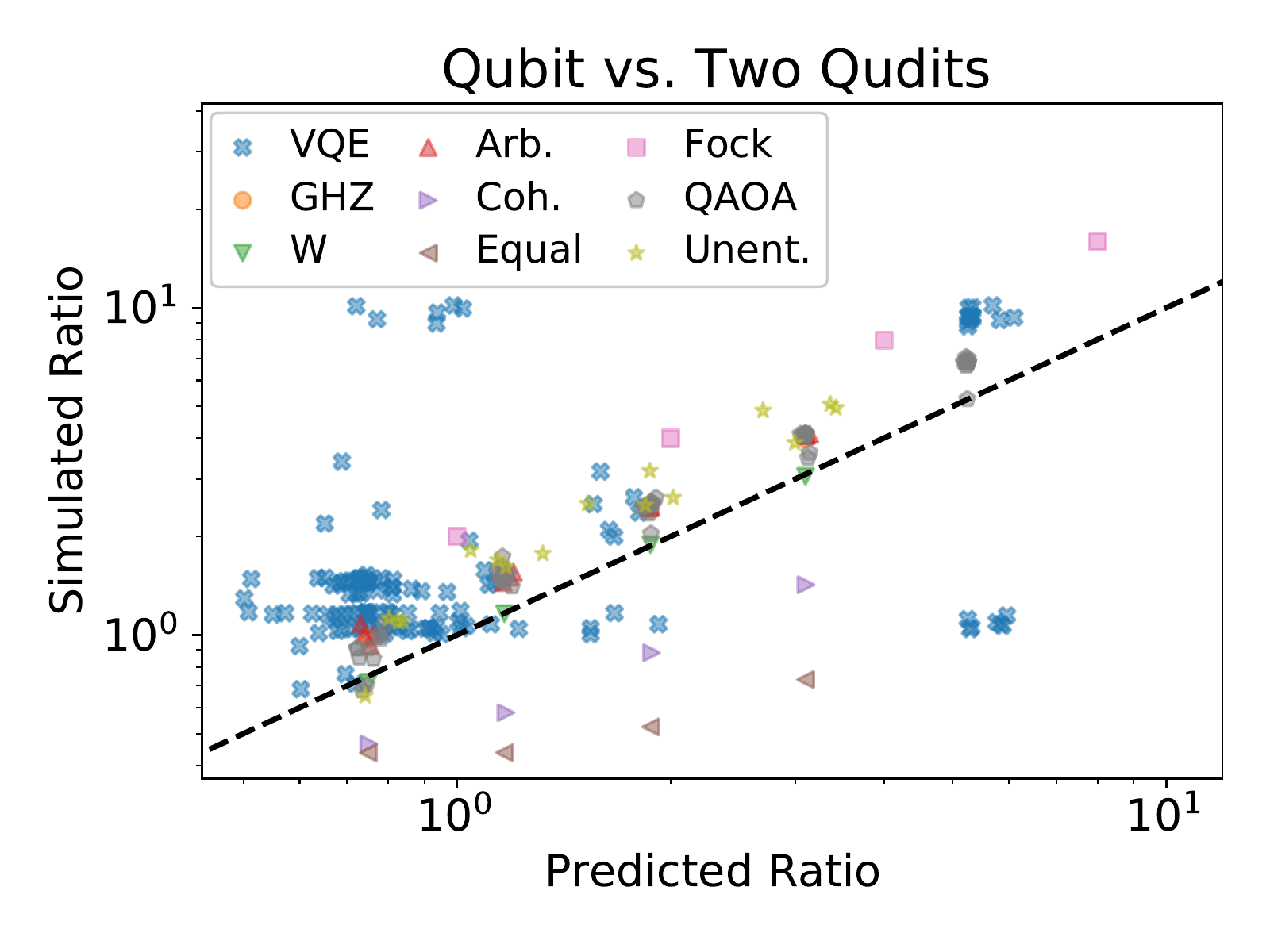}
  \caption{Comparison of the numerically simulated and analytically predicted scaling ratios between a
    qubit-based quantum memory with both amplitude damping and dephasing and an
    array of two qudits with only amplitude damping, for many quantum states.
    The dashed black line is
    $y=x$; the closer the points are to this line, the better the
    prediction.}\label{fig:int_qudit}
\end{figure}

Using either the formal derivation or intuitive arguments, one can easily generate ratios for the
performance of other quantum memories. For instance,
another potential architecture for a quantum memory is an array of qudits of
some intermediate size between a qubit ($d=2$) and a single qudit
(\change{$d=2^{n_q}$}). For illustrative purposes, we choose an array of two qudits
with \change{$d=2^{n_q-1}$}. The predicted ratio is, then,
\begin{equation}\label{eqn:ratio_int}
\frac{t_b}{t_{\mathrm{int}}} \approx \frac{\langle n_{\mathrm{int}} \rangle}{2\langle n_b \rangle},
\end{equation}
where the subscript $\mathrm{int}$ denotes the ``intermediate'' qudit system,
$\langle n_{\mathrm{int}} \rangle = \langle  \sum_i c_i^\dagger c_i\rangle$,
and $c_i$ is the \change{annihilation} operator for the intermediate qudit. Similar
comparisons between the intermediate qudit-based quantum memory and single
qudit-based quantum memory can also be performed. Figure~\ref{fig:int_qudit}
shows the comparison of the simulated and predicted scaling ratios between a
qubit-based quantum memory with both amplitude damping and dephasing and the
intermediate qudit-based quantum memory with only amplitude damping. Similar to
the comparison to the single qudit-based quantum memory (see
Fig.~\ref{fig:ratio_pred}), the predicted and simulated ratios are in good
agreement. The magnitude of the ratios, both simulated and predicted, are
considerably smaller since the average number of excitations in the intermediate
qudit-based quantum memory ($\langle n_{\mathrm{int}} \rangle$) is considerably
smaller than that in the single qudit-based quantum memory ($\langle n_d
\rangle$). Approximate performance ratios for other noise models can also be
generated within this framework. For example, in many qubit-based quantum
devices, the noise rates vary between qubits~\cite{tannu2018case} or even in
time~\cite{kandala2019error}. In such a disordered system, our assumption of
an equal noise rate, $\gamma$, for all channels breaks down. In this case,
instead of using the overall average number of excitations, $\langle n_b
\rangle$, the average excitation per qubit needs to be weighted by its overall
noise contribution, giving
\begin{equation}
  \sqrt{\mathcal{F}_t} = \sum_k \sum_j\frac{(-\gamma_j t_{dis})^k}{k!} \langle n_j^k \rangle,
\end{equation}
where $\langle  n_j^k \rangle =  \big|\langle \psi(0) | \big(\sigma_j^\dagger \sigma_j\big)^k | \psi(0)
\rangle\big|$ and we have assumed that the dephasing and amplitude-damping rates
are the same for a given qubit but different between qubits. Comparing an
ordered qubit register with a disordered qubit
register, we find that the approximate ratio of times to reach a target fidelity
is, to first order,
\begin{equation}
  \frac{t_b}{t_{dis}} = \frac{\sum_j \gamma_j \langle n_j \rangle}{\gamma \langle n_b \rangle}.
\end{equation}

Architectures with other noise channels, beyond amplitude damping and dephasing,
can also be included as long as they can be written in the non-Hermitian
formalism. There is no guarantee that every noise channel will lead to as simple
 a formula as Eq.~\eqref{eqn:ratio_prediction}, but many will because of the
simple relationship of the fidelity to the moments of the operators. A two-qubit
correlated amplitude damping channel, for instance, would give a contribution of
$\langle  n_{b,i} n_{b,j}\rangle$ (in addition to any single-qubit noise).

\begin{figure}
  \includegraphics[width=\columnwidth]{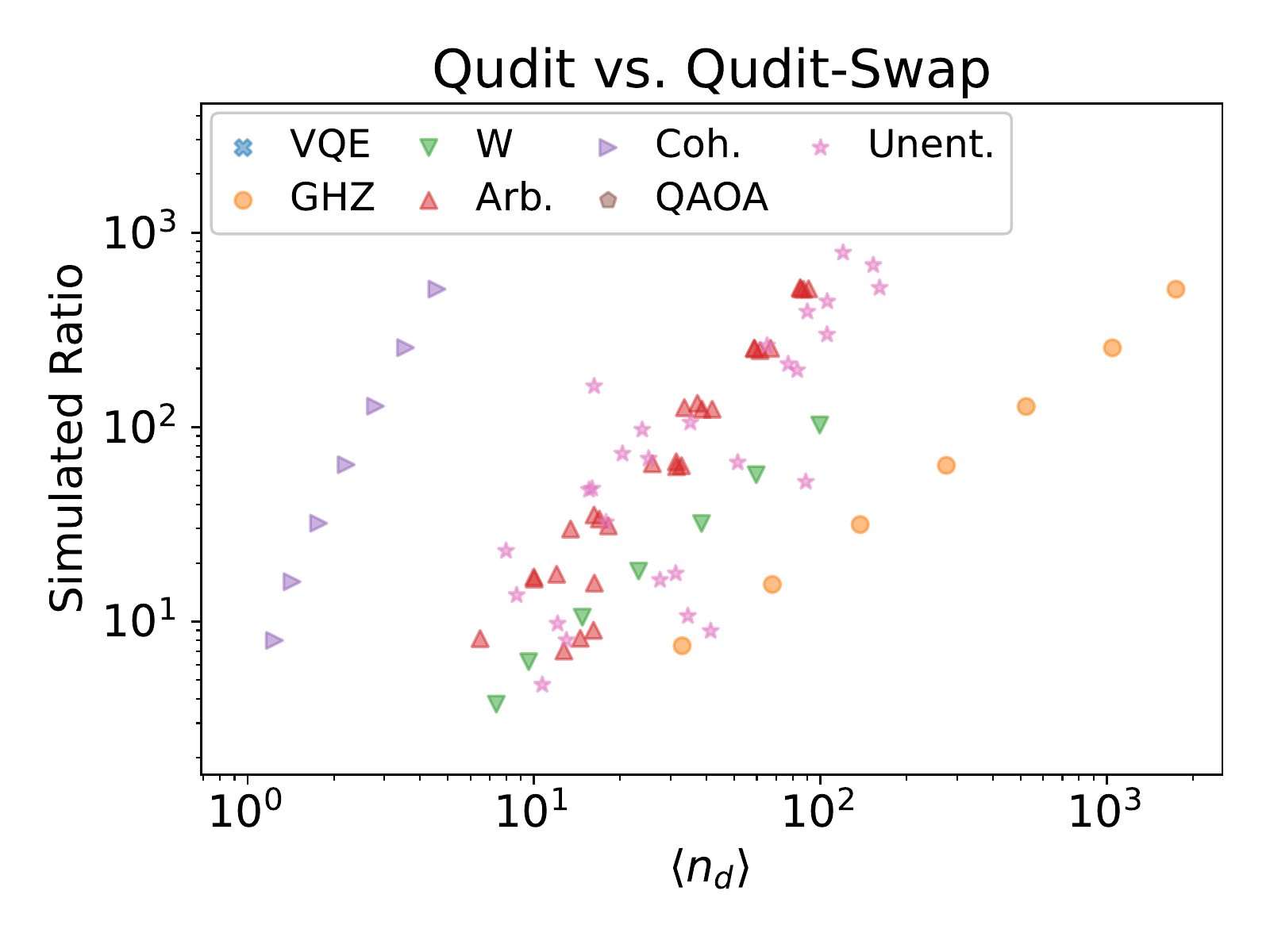}
  \caption{Performance enhancement from reordering the quantum information in a qudit-based
  quantum memory with amplitude damping. The simulated performance gain grows with increasing average
  number of excitations in the original state.}\label{fig:swap}
\end{figure}

Our analysis of the correlation of the noise model
and the structure of the quantum information can help provide insights into ways
to potentially extend the lifetime of quantum states stored in quantum memories.
The overall noise, to first order, goes as the number of excitations in the system
(regardless of the architecture). Storing the quantum information such that the
largest amplitudes are in the lowest possible states will greatly reduce the
overall noise and thus increase the effective lifetime of the quantum
information. To demonstrate this benefit, we numerically sorted the amplitudes of the
quantum states and simulated the performance of a qudit-based quantum memory
with only amplitude damping with both the unsorted and sorted quantum states.
The comparison between the two layouts is shown in Fig.~\ref{fig:swap}. Reordering
states with a small number of amplitudes that happen to be in highly excited
states, such as the GHZ and W states, provides a large performance enhancement that
grows with system size. Random arbitrary and unentangled states see a modest
performance enhancement. The coherent state, on the other hand, actually
performs worse after reordering. After reordering, the coherent state is no
longer an eigenstate of the annihilation operator and thus loses its superior performance.
Reordering the amplitudes of an arbitrary
quantum state is generally very expensive; a quantum sort of $n$ items, for
example, will take at least $\mathcal{O}\big(n \log (n)\big)$ steps~\cite{hoyer2001quantum}. However, in
the case where classical information is being loaded onto a quantum device
through, say, a QRAM technique~\cite{Giovannetti2008}, carefully arranging the amplitudes can
increase the performance of the quantum memory, especially in a qudit-based
system, where the decay of high \change{Fock} states is significantly larger
than lower \change{Fock} states. Techniques, such as reordering, based on the
correlation of the structure of the quantum state with the noise model can be
used in addition to quantum error
correction~\cite{gottesman1997stabilizer,leghtas2013hardware,raussendorf2007fault,fowler2012surface}
and other error mitigation
techniques~\cite{kandala2019error,otten2019accounting,otten2019recovering}.

\begin{table}[t]
  \centering
  \begin{ruledtabular}
    \begin{tabular}{ccc}
      & Simulated Ratio & Predicted Ratio \\
      \hline
      Coherent & 0.96 & 51.29\\
      GHZ & 51.67 & 51.15\\
      W & 44.62 & 51.15\\
      Equal & 8.92 & 51.15\\
      Fock & 366.00 & 256.00\\
      VQE & 19.05 $\pm$ 2.04 & 14.41 $\pm$ 3.88\\
      QAOA & 49.44 $\pm$ 3.12 & 50.72 $\pm$ 0.43\\
      Arbitrary  & 49.08 $\pm$ 0.14 & 51.53 $\pm$ 0.38\\
      Unentangled  & 86.15 $\pm$ 20.89 & 58.93 $\pm$ 9.32\\
    \end{tabular}
  \end{ruledtabular}
  \caption{Simulated and predicted performance ratios between a qubit-based
    quantum memory with both amplitude damping and pure dephasing and a
    qudit-based quantum memory with only amplitude damping, for a variety of
    states of size $2^{10}$.}
  \label{tab:state_comparison}
\end{table}

 The derived
performance ratio, Eq.~\eqref{eqn:ratio_prediction}, is an experimentally
accessible quantity, even for unknown quantum states, and can be used to
understand the relative performance and engineering requirements between quantum
architectures. For some given quantum
state, choosing between two candidate quantum memory architectures would involve
measuring the average number of excitations in either device, as well as knowledge of
underlying noise models of each quantum memory.
Measuring the average
number of excitations is a simple experimental technique that can be done with
$\mathcal{O}(1)$ experiments, in both qubits (through standard single-qubit
measurements~\cite{naghiloo2019introduction}) and qudits (through multilevel quantum tomography techniques~\cite{chakram2020seamless, chakram2020multimode} or other means), as
long as many copies of the quantum state are available. The relative performance
between the two quantum memories, assuming equivalent noise rates, could then be
predicted by using a formula like Eq.~\eqref{eqn:ratio_prediction}. The relative
performance could then be compared with the relative noise rates to understand
which quantum architecture would perform better. As an explicit example,
we return to the qubit-based quantum memory with amplitude damping and pure
dephasing and the qudit-based quantum memory with only amplitude damping. For
one instance of an arbitrary state of size $2^4$ with random wavefunction
coefficients, we find that $\langle n_d \rangle = 8.16$ and $\langle n_b
\rangle = 2.08$, leading to a performance ratio assuming equivalent noise
rates, according to Eq.~\eqref{eqn:ratio_prediction}, of $\approx 1.96$ (the
simulated performance ratio for target fidelity $\mathcal{F}_t=0.75$ is 2.44);
the qubit-based quantum memory will have reliably stored the quantum
state for about twice as long as the qudit-based quantum memory. Put another way,
the qudit-based quantum memory would need to have half the noise rate in order to perform
as well as the qubit-based quantum memory. Similar comparisons can be made for
other quantum states. Table~\ref{tab:state_comparison} shows the predicted and
simulated performance ratios for a variety of quantum states. We find
performance ratios on the order of 10--100 for most states of size $2^{10}$, with
the coherent state and Fock states being strong outliers.
3D cavity qudits can have $T_1$ times
that are  more than 100$\times$ longer than the $T_1$ times of transmon qubits~\cite{Reagor2016, Romanenko_3d}
Thus, for states of
size  $2^{10}\approx 1000$, 3D cavity architectures will likely perform
better than qubit-based systems. Above that, a qubit-based system will perform
better.

\section{Concluding Remarks}
We presented a detailed study of the interplay between the structure of quantum information
and the physical noise models of various quantum memories. We demonstrated simple and
experimentally relevant ways
of comparing the expected performance of quantum memory architectures for
specific quantum states. Although we focused primarily  on two paradigmatic
devices, superconducting qubits with both amplitude
damping and dephasing channels and superconducting cavities with only
amplitude damping, our methods can easily be extended
to other quantum systems such as trapped ions and photonic systems.
We utilized both numerical simulations using the Lindblad master equation and an approximate
non-Hermitian formalism to analyze the behavior of a wide variety of interesting and
useful quantum states. Our approximate non-Hermitian analysis provides a simple formula that gives
intuition on the performance of a given quantum state
stored in different devices. As a practical example of the application of our method, we
demonstrated that the superconducting cavities are  viable candidates
for quantum memories up to around 1,000 levels for many classes of states because of
their significantly longer $T_1$ times.
Beyond that, the increased decay from the higher levels lowers the overall
fidelity for many of the states,
and an array of superconducting qubits becomes a more viable quantum memory. Furthermore, we showed that reducing the total number of excitations in the mapping of data to a quantum state
can help increase the overall lifetime of the state and thus provides a way, through state
engineering, to increase the performance of a quantum memory device.
Our method, given its simplicity and experimental relevance, could be used as part of a heuristic for
a quantum compiler for hybrid quantum devices. As long as multiple copies of a state can be created, a
simple interrogation of the state, measuring the number
of excitations when stored in the various possible subcomponents of the overall device, can be used to
decide where to store a state. As the complexity of hybrid quantum devices grows, simple heuristic methods
for understanding the performance of each subcomponent, such as the method we present here, will be important
for maximizing the overall performance of the device and can help in the initial design.

\section*{Acknowledgments}
This work is supported by the U.S.\ Department of Energy, Office of Science, Office of High Energy Physics through a QuantISED program grant:  Large Scale Simulations of Quantum Systems on HPC with Analytics for HEP Algorithms (0000246788). This manuscript has been authored by Fermi Research Alliance, LLC under Contract No.\ DE-AC02-07CH11359 with the U.S. Department of Energy, Office of Science, Office of High Energy Physics. We gratefully
acknowledge the computing resources provided on Bebop, a high-performance computing cluster operated by
the Laboratory Computing Resource Center at Argonne
National Laboratory.
Argonne National Laboratory's work was supported by the U.S. Department of Energy, Office of Science and Technology, under contract DE-AC02-06CH11357.
 We also thank K. B. Whaley for useful discussion.

\appendix

\section{\label{ap:A}Derivation of Approximate Solution}

To derive the approximate solutions of Eq.~\eqref{eqn:qd_fid} and
Eq.~\eqref{eqn:qb_fid}, we begin with the generic solution to the
non-Hermitian Schr\"{o}dinger equation of Eq.~\eqref{eqn:nh} with initial condition
$|\psi(0)\rangle$,
\begin{equation}\label{eqn:sch_sol}
  |\psi(t) \rangle = e^{-\sum_j \frac{\gamma_j}{2} C_j^\dagger C_j } |\psi(0)\rangle.
\end{equation}

We seek to understand the evolution of the fidelity with respect to the initial
state,
\begin{equation}\label{eqn:fidelity}
  F(t) = |\langle \psi(0) | \psi(t) \rangle|^2.
\end{equation}
We derive the following using the square root of the fidelity,
$\sqrt{F(t)}$, because of the increased ease of typesetting. We can
expand the fidelity using the solution to Schr\"{o}dinger equation,
Eq.~\ref{eqn:sch_sol},
\begin{equation}\label{eqn:sqrt_fid2}
  \sqrt{F(t)} = |\langle \psi(0) | e^{-\sum_j \frac{\gamma_j}{2} C_j^\dagger C_j } | \psi(0) \rangle|.
\end{equation}
The equation is valid for any initial condition, $|\psi(0)\rangle$ and
any system with any noise operators, $C_i$. We will now derive specific formulas
for several different noise models.

\subsection{Single Qudit with Amplitude Damping}
A single qudit with amplitude damping has only a single noise operator, the
\change{annihiliation} operator for an $n$-level system, $b$. We will also expand the initial
state in terms of the basis states of the qudit, leading to
\begin{equation}\label{eqn:qd_sqrt_fid}
  \sqrt{F(t)} = \sum_{k,j}|\langle j | \alpha_j^* \alpha_k  e^{-\frac{\gamma}{2} b^\dagger b t} | k \rangle|.
\end{equation}

To simplify this expression,
we expand the exponential
\begin{equation}\label{eqn:qd_expansion}
  e^{-\frac{\gamma}{2} b^\dagger b t} | k \rangle = \sum_l \frac{(-\frac{\gamma t}{2})^l(b^\dagger b)^l}{l!} | k \rangle.
\end{equation}
Because $b^\dagger b |k\rangle = k |k\rangle$, we can rewrite this equation
as
\begin{equation}\label{eqn:qd_simplification}
  e^{-\frac{\gamma}{2} b^\dagger b t  } | k \rangle = \sum_l \frac{(-\frac{\gamma t}{2})^l(k)^l}{l!} | k \rangle = e^{-\frac{\gamma}{2} k t} | k \rangle,
\end{equation}
which removes the operator from the exponential. We can now use this
simplification
in the fidelity expression,
Eq.~\eqref{eqn:qd_sqrt_fid}.  Combining this with the orthonormality of the
basis states, we have
\begin{equation}
  \sqrt{F(t)} = \sum_{j}|\alpha_j|^2 e^{-\frac{\gamma}{2} jt},
\end{equation}
which is the equation in the main text, Eq.~\eqref{eqn:qd_fid}.

\subsection{Qubits with Amplitude Damping and Dephasing}
The derivation for other noise models, such as a qubit register with both
amplitude damping and dephasing, follows the same logic. The primary difference
is the noise operators, $C_i$, and how they act on the basis states, $|j\rangle$.
For example, on a qubit register, the amplitude damping noise channels are a sum
of operators, $\sum_i \sigma_i^\dagger \sigma_i$, where $\sigma$ is the \change{annihilation}
operator for a two-level system. The expansion of the exponential in
Eq.~\eqref{eqn:qd_expansion} changes. The basis states, $|j\rangle$, are now
bit strings of length the number of qubits, $n_q$. If qubit $i$ is in its
excited state, $|1\rangle$, the action of its noise operator $\sigma_i^\dagger
\sigma_i$ will return 1; otherwise, it will return 0. The sum of the action of
all  the amplitude damping noise channels is then  just a count of the
number of excited qudits in the basis state. This number is known as the Hamming
weight, $w(j)$. The dephasing operator, by similar arguments, contributes a
term proportional to the Hamming weight. Together, this gives the equation in
the main text, Eq.~\eqref{eqn:qb_fid}.

\section{\label{ap:ftar} Changing $\mathcal{F}_t$}
\begin{figure*}
  \centering
  \subfloat[$\mathcal{F}_t = 0.7$]{\includegraphics[width=0.97\columnwidth]{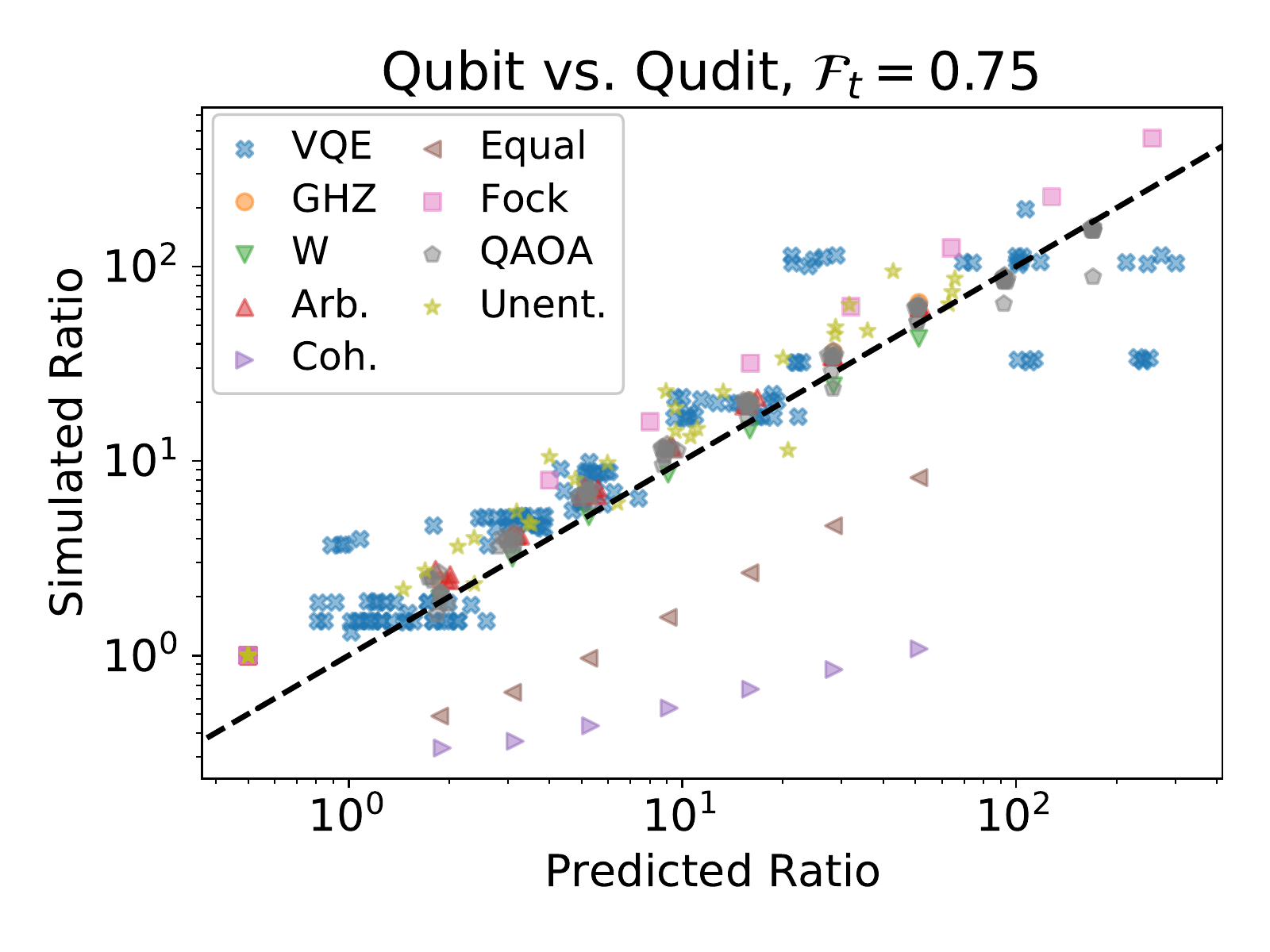}}\quad
  \subfloat[$\mathcal{F}_t = 0.9$]{\includegraphics[width=0.97\columnwidth]{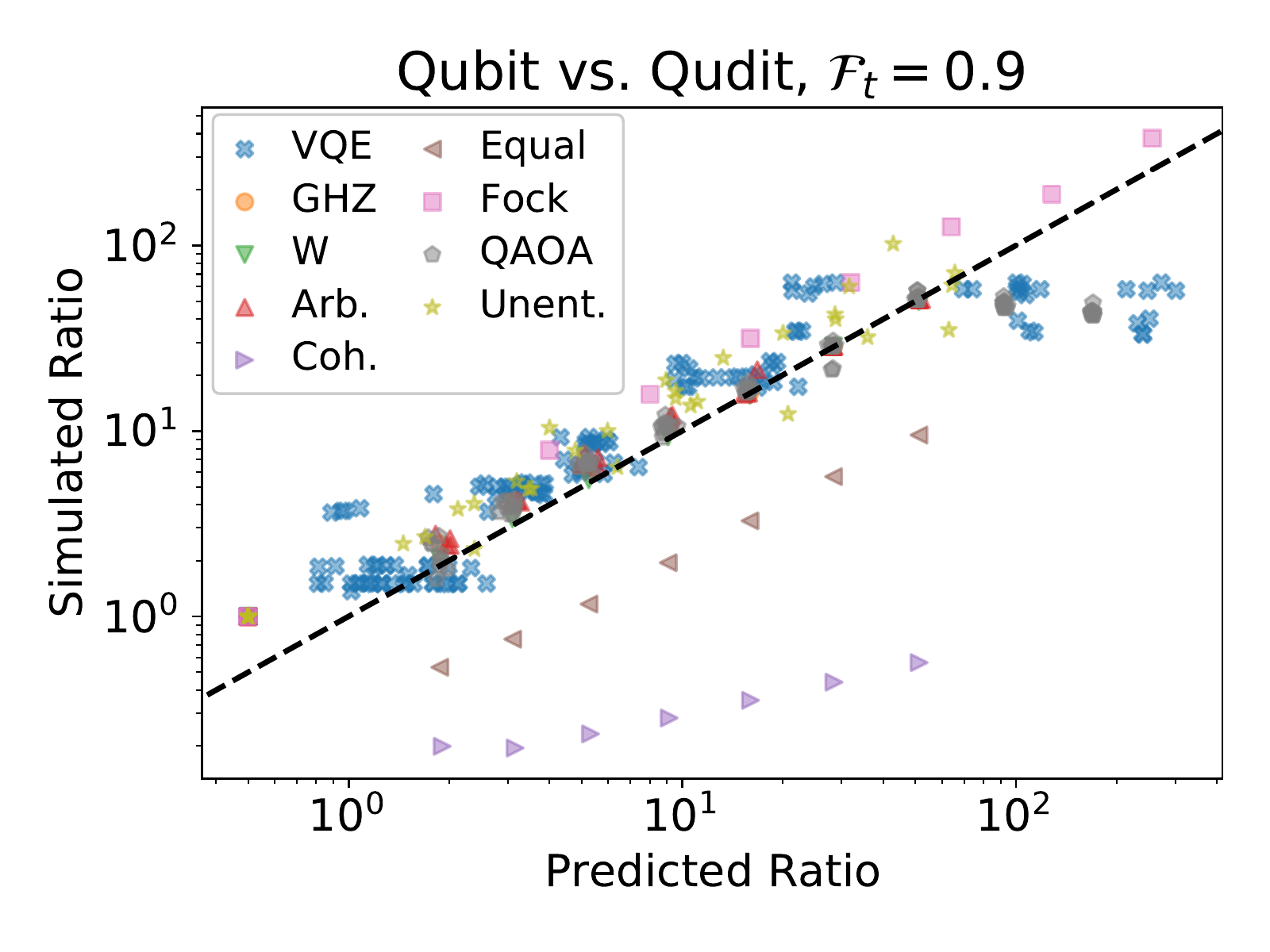}}
  \caption{Comparison of the simulated and predicted scaling ratios between a
    qubit-based quantum memory with both amplitude damping and dephasing and a
    qudit-based quantum memory with only amplitude damping, for a wide of variety
    of interesting quantum states. The two different target fidelities here (in
    addition to Fig.~\ref{fig:ratio_pred}) show that the specific target
    fidelity does not affect the overall conclusions.} \label{fig:scaling_diff_ftar}
\end{figure*}

\begin{figure}
  \centering
  \includegraphics[width=0.97\columnwidth]{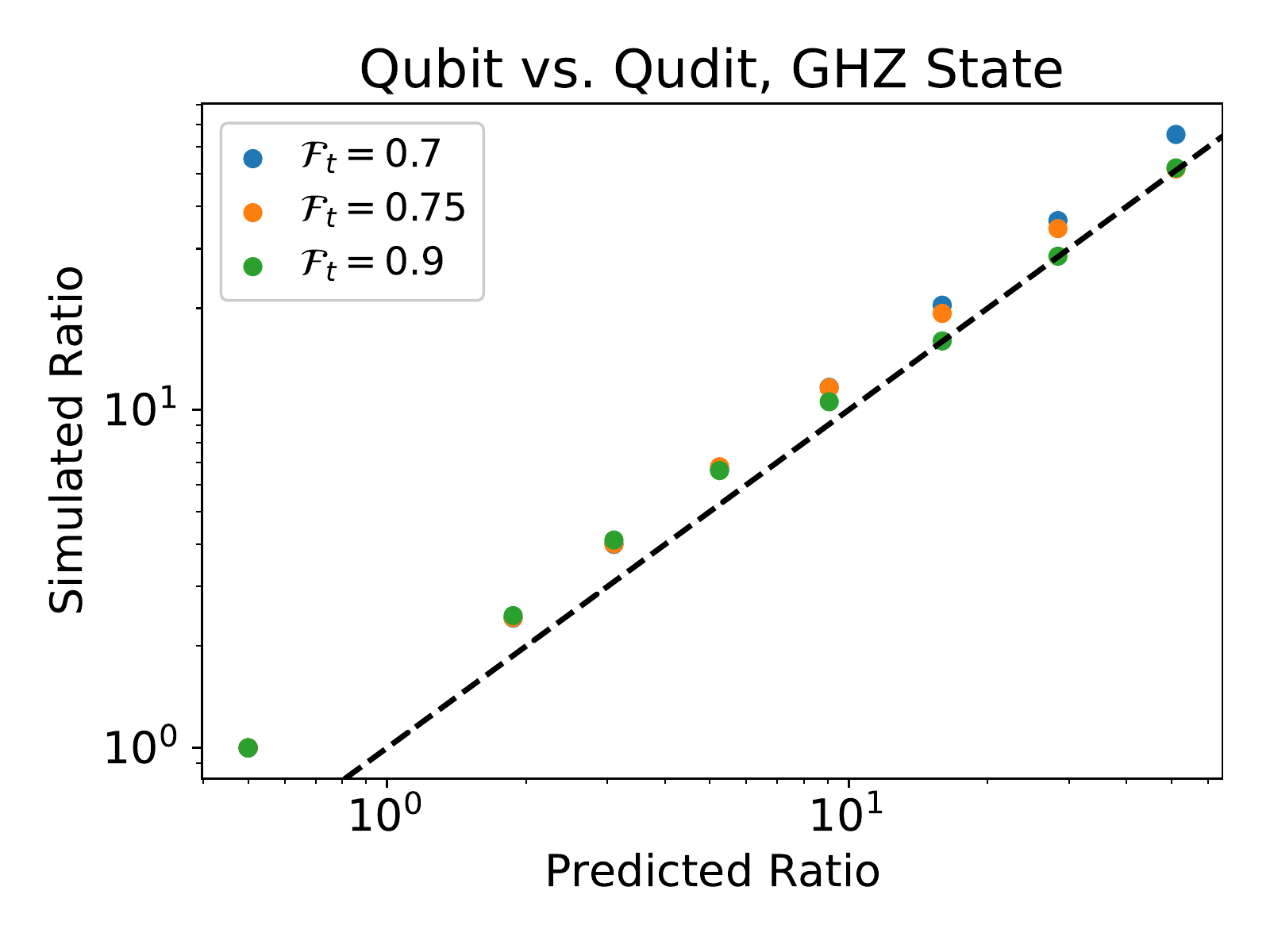}
  \caption{Comparison of the simulated and predicted scaling ratios between a
    qubit-based quantum memory with both amplitude damping and dephasing and a
    qudit-based quantum memory with only amplitude damping for the GHZ state at
    three different target fidelities. While the exact location of the simulated
  scaling changes, the values are reasonably close.} \label{fig:ghz_ftar}
\end{figure}

Figure~\ref{fig:scaling_diff_ftar} shows the results of applying the analysis of
the main text (that is, using the predicted scaling ratio of
Eq.~\eqref{eqn:ratio_prediction}) to the various quantum states studied at two
target fidelities ($\mathcal{F}_t = 0.75$ and 0.9). These results, in
addition to the results of Fig.~\ref{fig:ratio_pred}, all support the efficacy
of the scaling prediction. While the predicted ratio is insensitive to the
target fidelity, the simulated ratio is sensitive to it. Figure~\ref{fig:ghz_ftar} shows the
difference between simulated ratios for the GHZ state. The exact position of the
simulated ratio changes for GHZ states with larger numbers of excitations in the
qudit, but the differences are negligible on the studied scale.

\section{\label{ap:states}Description of Quantum States}
In this section, we describe all  the states used in the main text. Unless
mentioned otherwise, these states were generated by using utilities available within
the QuTiP~\cite{johansson_qutip:_2012} package.

\textbf{GHZ State}
The GHZ state is defined in Eq.~\eqref{eq:ghz_state}.

\textbf{W State}
The W state for $n_q$ qubits is defined as the equal superposition of all states,
where one qubit is excited ($|1\rangle$) and all other qubits are in their
ground state ($|0\rangle$).

\textbf{Equal Superposition State}
The equal superposition state is defined as the state with an equal
superposition of all possible basis states.

\textbf{Fock States}
A Fock state, or number state, is defined as a state with a specific number of
excitations. In this work, we use the Fock states $|8\rangle,$ $|16\rangle,$
$|32\rangle,$ $|64\rangle,$ $|128\rangle,$ $|256\rangle,$ and $|512\rangle$, when
represented as a qudit state.

\textbf{Coherent States}
A coherent state is generally defined as
\begin{equation}\label{eq:coherent_state}
  |\alpha \rangle = e^{-|\alpha|^2/2} \sum_{n=0}^{\infty}
  \frac{\alpha^n}{\sqrt{n!}}|n\rangle,
\end{equation}
where $\alpha$ is generally a complex number. In this work, however, we focus on
quantum systems that have a limit on the number of excitations that can be in
the system. We instead use the follow definition,
\begin{equation}\label{eq:disp}
  |\alpha \rangle =  e^{\alpha (a^\dagger -  a)} | 0 \rangle,
\end{equation}
where $a$ is a truncated \change{annihilation} operator. This definition of the coherent state gives slightly different
amplitudes from those obtained with the analytic formula of Eq.~\eqref{eq:coherent_state},
especially in the small truncation limit. We use $\alpha = \sqrt{n_t/2}$, where
$n_t$ is the maximum number of excitations allowed in the qudit register  and
$n_t = $ 16, 32, 64, 128, 256, 512, and 1024.

\textbf{Chemical States}
To generate states relevant to quantum chemistry studies, we use Qiskit's
Aqua~\cite{Qiskit} package using parity mapping to map spin orbitals to
qubits~\cite{bravyi2017tapering} for all molecules. Rather than solve a
variational quantum eigensolver instance, we instead use exact diagonalization
to find the exact ground and first excited states for all molecules. We generate
states for H$_2$ with and without two-qubit reduction~\cite{bravyi2017tapering},
LiH with two-qubit reduction with and without a frozen
core~\cite{kandala-nature-2017}, H$_4$ with and without two-qubit reduction, and
H$_2$O with two-qubit reduction. These states span Hilbert space sizes from 4 to
4,096.

\textbf{QAOA States}
To generate quantum approximate optimization algorithm (QAOA) states, we
generate Erdős-Rényi graphs~\cite{erdHos1960evolution} of size $n$ with
probability $0.5$ of creating an edge between any two nodes. We then use the QAOA
solver within Qiskit~\cite{Qiskit} with $p=4$ steps to solve for the MaxCut of
the graph~\cite{shaydulin2019network}. We generate and solve
ten graphs each of size $n=$ 4, 5, 6, 7, 8, 9, 10, 11, and 12.

\textbf{Arbitrary States}
We randomly generate arbitrary states by creating dense vectors of uniform random numbers in
the range $[-0.5,0.5]$ for both the real and imaginary parts and then normalize
the vectors. We create four random states for each total Hilbert space size of 16,
32, 64, 128, 256, 512, and 1024.

\textbf{Unentangled States}
We randomly generate unentangled- states as described above for two-level systems and then take the
tensor products of several such qubit states to create random, unentangled
states. We generate four random states for tensor products of size 4, 5, 6, 7,
8, 9, and 10 qubits.

\section{\label{ap:comparison}Comparison of Numerical and Analytic Dynamics}
\begin{figure}
  \centering
  \includegraphics[width=0.97\columnwidth]{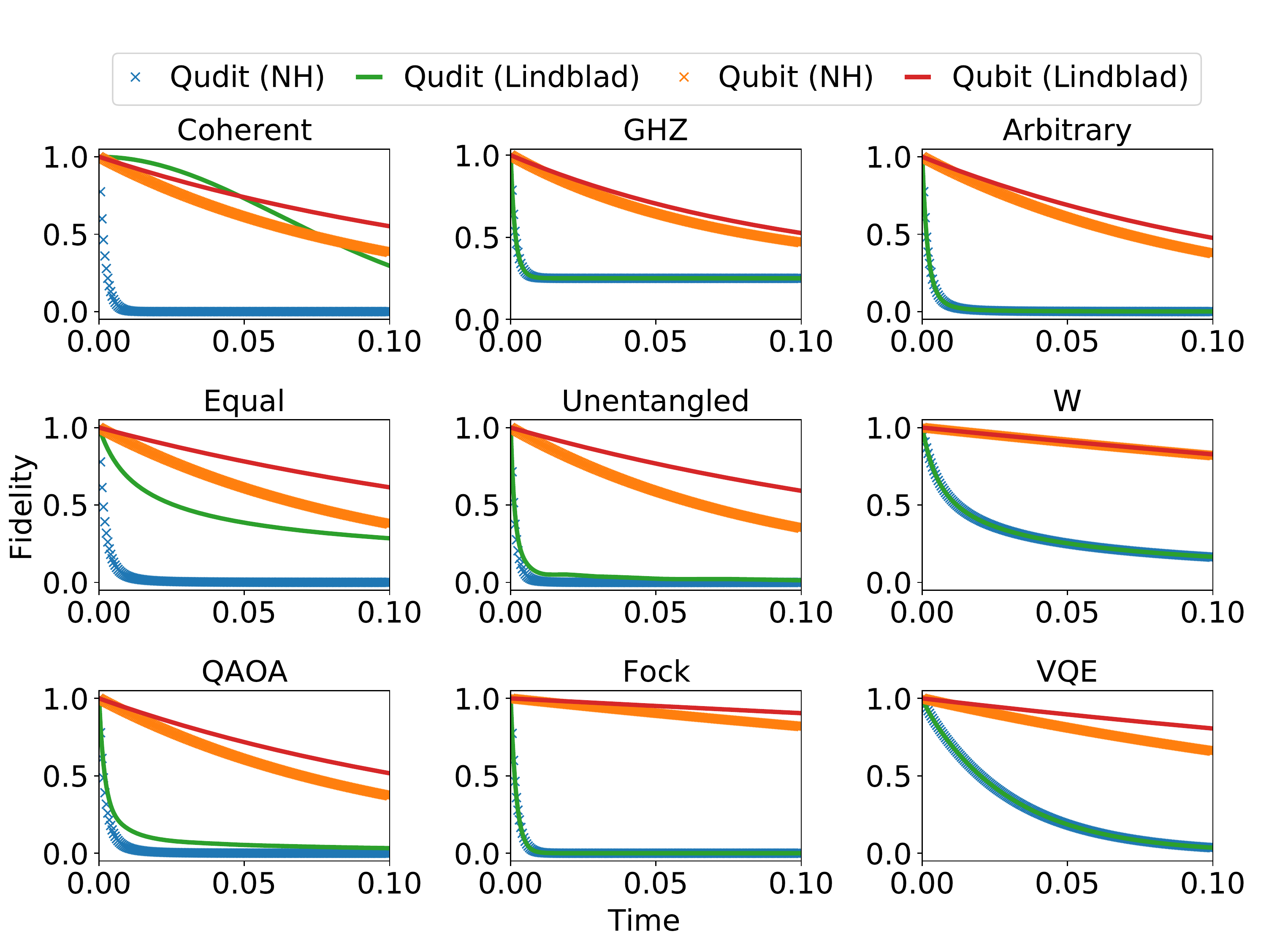}
  \caption{Comparison of the full Lindblad dynamics and the non-Hermitian (NH) approximation
  for specific $n_q=10$ instances of the various states studied.} \label{fig:comp_formalisms}
\end{figure}

\change{We compare the full Lindblad dynamics of both the qubit-based (Eq.~\eqref{eqn:qb_lme}) and
qudit-based (Eq.~\eqref{eqn:qd_lme}) quantum memories with their respective non-Hermitian dynamics
in Fig.~\ref{fig:comp_formalisms}. We find that the non-Hermitian (NH) dynamics, at least for the
short-times we are interested in, provides a good approximation for many of the states.
For some states, such as the coherent and equal superposition states, the difference in
approximation error between the qubit and qudit models is stark. For example, in the coherent
state, the qubit system sees
good agreement between the full Lindblad dynamics and the approximate non-Hermitian dynamics,
but the non-Hermitian dynamics greatly underestimates the true Lindblad dynamics for the qudit system.
Correspondingly, this leads to the large errors seen for some of the states in the predicted
ratios (see Table~\ref{tab:state_comparison}).}

\bibliography{qudit_bit_paper}
\end{document}